\documentclass[9.5pt,journal,compsoc]{IEEEtran}

\ifCLASSOPTIONcompsoc
  \usepackage[nocompress]{cite}
\else
  \usepackage{cite}
\fi
\usepackage[pdftex]{graphicx}
\usepackage{array}

\usepackage{subfig}
\usepackage{url}

\begin{document}
%
\title{Machine Learning based Anomaly Detection for 5G Networks}
%
%
%
%

\author{Jordan~Lam,~\IEEEmembership{Macquarie University,}
        Robert~Abbas,~\IEEEmembership{Macquarie University}
\IEEEcompsocitemizethanks{\IEEEcompsocthanksitem Jordan Lam - Macquarie University.\protect\\
E-mail: jordan.lam@students.mq.edu.au}
\IEEEcompsocitemizethanks{\IEEEcompsocthanksitem Robert Abbas - Macquarie University.\protect\\
E-mail: robert.abbas@mq.edu.au}

\thanks{Last revised March 6th, 2020.}}

\IEEEtitleabstractindextext{%
\begin{abstract}
 Protecting the networks of tomorrow is set to be a challenging domain due to increasing cyber security threats and widening attack surfaces created by the Internet of Things (IoT), increased network heterogeneity, increased use of virtualisation technologies and distributed architectures. This paper proposes SDS (Software Defined Security) as a means to provide an automated, flexible and scalable network defence system. SDS will harness current advances in machine learning to design a CNN (Convolutional Neural Network) using NAS (Neural Architecture Search) to detect anomalous network traffic. SDS can be applied to an intrusion detection system to create a more proactive and end-to-end defence for a 5G network. To test this assumption, normal and anomalous network flows from a simulated environment have been collected and analyzed with a CNN. The results from this method are promising as the model has identified benign traffic with a 100\% accuracy rate and anomalous traffic with a 96.4\% detection rate. This demonstrates the effectiveness of network flow analysis for a variety of common malicious attacks and also provides a viable option for detection of encrypted malicious network traffic.
\end{abstract}

}

\maketitle

\IEEEdisplaynontitleabstractindextext

%
\IEEEpeerreviewmaketitle

\IEEEraisesectionheading{\section{Introduction}\label{sec:introduction}}

%
%
%
%
\IEEEPARstart {O}{ver} the last decade, exponential increases in computing power has allowed machine learning models such as neural networks to operate with greater efficiency and deliver increasingly accurate results. This in turn has led to many novel applications of machine learning to be conceived from traditional areas of research such as speech recognition and computer vision. In this paper one such novel application will be investigated, the application of a CNN to analyse network traffic with the goal of providing an adaptive security solution for the diverse threat landscape of 5G networks. This application will be implemented by collecting benign and anomalous network flow data from a simulated environment and using these flows as the input data for a CNN. An anomalous network flow can be defined as behaviour that is unusual or does not fit with regular traffic patterns for a particular user, business or entity. This paper will assume anomalous network flows as malicious for testing purposes, however in a real world scenario anomalous traffic may not be malicious but is still worthy for analysis due to potential future business impacts.

The layout of this paper is as follows, firstly the 5G security landscape will be investigated, this includes examining the current environment in relation to the security architectures that 5G is able to inherit from LTE (Long Term Evolution) and current 3GPP (3rd Generation Partnership Project) developments in 5G security. Future security concerns for 5G networks will then be examined, these include how the exponentially growing number of IoT devices is changing the security landscape, managing multiple technologies, increased virtualisation threats, managing distributed architectures and network slices. A solution will then be proposed through the implementation of a SDS system which utilises machine learning to a 5G network. The system is designed to access traffic from both the backhaul link into the core network and from the interconnect link out of the core network to detect end-to-end threats and actively update appropriate security policies. 

Secondly the applications of machine learning will be investigated in terms of current advances in anomaly detection. These current advances in this area of research will be discussed and applications of neural network architecture will be compared to demonstrate the benefits of CNN's. Further discussion will include assessing the design of a CNN with NAS and the application of autoML (Automated Machine Learning) to analyse the provided data set to produce a specific CNN model layout. Finally the layout of the collected data set will be examined and the data will be pre-processed into images which are acceptable to the model. Results will then be collected from applying the model (Appendix A) to the data set and these results will be evaluated for their viability and application to 5G and IoT security use cases. The goal of this project is to provide some insight into the effectiveness of machine learning in intrusion detection applications and show how this solution can be completely defined through software allowing greater flexibility, scalabililty and portability in a 5G network.


 


\section{5G Network Security Environment}
Environmental needs and threat models are changing as malicious actors become more advanced and networks become more complex and heterogeneous. IoT devices are also predicted to increase from the current number of 27 billion devices to 75 billion by 2025, this is a further cause for concern in ensuring these devices can not be used in attacks against mobile networks \cite{iot}. Historically mobile networks have been built with security in mind from the ground up, utilising multiple defences implemented in all layers of the network. This is a good sign for the design of future 5G networks, however networks will become increasingly heterogeneous as legacy, LTE and 5G network traffic have to be supported simultaneously and have increased reliance on software based and virtualisation technologies. The large difference in 5G networks in comparison to LTE networks will bring much greater security risks and is a cause for concern for network operators in maintaining a secure, stable and reliable service.

\subsubsection{Current Threat Landscape}
Telecommunication networks can be broken down to include four major logical elements these are the radio access network, core network, transport network and inter-connect network \cite{ericsson}. Each of these network elements is comprised of three planes which are each responsible for carrying different types of traffic. A graphical overview of how these elements interact is shown in Figure 1.

\vspace{0.5cm}
\hspace*{-0.5cm} 
\makebox[\textwidth][l]{\includegraphics[scale=0.3,width= 0.48\textwidth]{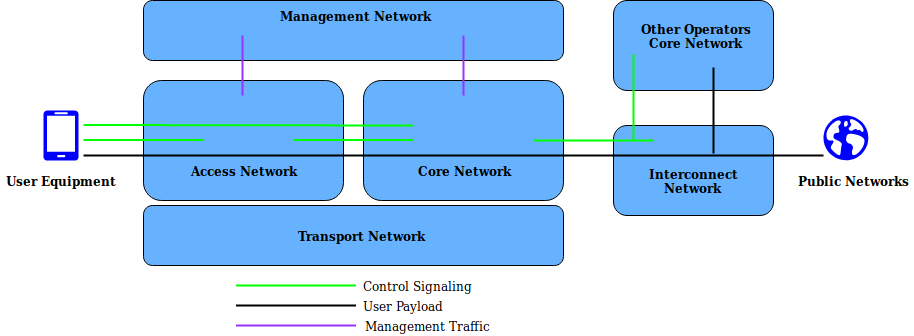}}
\textit{Fig. 1: Big Picture: Telecommunication Networks}

These are defined as the control plane which carries signalling traffic, the user plane which carries the payload (actual traffic) and the management plane which carries the administrative traffic \cite{ericsson}. From a security perspective all three of these planes are exposed to unique threats and also uniform threats which relation to security all three planes can be exposed to unique threats and there are also uniform threats which can affect all three planes simultaneously. Network security is implemented into telecommunication networks in the following four phases \cite{ericsson}:

\begin{itemize}
    \item Standardisation: Operators, vendors and stakeholders set standards for how networks globally will operate. Standards are also defined in relation to protecting networks against any type of malicious actor.
    \item Network Design: Network vendors design, develop and implement the agreed standards into functional network elements and systems, ensuring the end product is both functional and secure.
    \item Network Configuration: During the network deployment phase, networks are configured to achieve a set security level, this is critical in setting security parameters and further strengthening both network security and resilience.
    \item Network Deployment and Operation: This is the operational phase of the network, achieving defined security levels is dependent on appropriate network deployment and operation.
\end{itemize}

In terms of 5G, this technology can be defined as not only providing another incremental upgrade in terms of speed and latency but rather an enabler of a new set of services and use cases, with the most unique selling point of 5G being the realisation of a true IoT and inter-networked environment that will impact all parts of society \cite{americas}. However the main factor that will determine whether or not 5G can live up to it's potential is the question of how secure and stable can 5G deliver these new services? Traffic sensors and Vehicle-to-infrastructure services are one use case of IoT devices \cite{americas} and it is critical that even these basic devices are protected as they are highly vulnerable to DDoS (Distributed Denial of Service) attacks. A clear example of compromised IoT security is the Mirai attack that managed to control 600,000 vulnerable IoT devices in a botnet, applying massive DDoS attacks on high profile services such as OVH and Dyn \cite{mirai}. 

Fortunately past telecommunication networks have ensured that security is a top architectural concern, which is good news for 5G. For example in relation to LTE security the 3GPP Release 8 added a variety of advanced security and authentication mechanisms via nodes such as the services capability server, while Release 11 provided additional capabilities to the core network for secure access \cite{americas}. These concerns of trust and authentication within the network also carry over to 5G networks as 3GPP Release 15 adds two mandatory authentication options for 5G and builds a trust model through key separation \cite{3gpp}. In this way LTE network security provides a foundation for enabling future 5G security processes. In terms of physical layer wireless security the telecommunications industry is held in high regard in comparison to other wireless technologies, even a mobile phone's use of licensed spectrum adds additional layers of security to aid in preventing against eavesdropping on data, voice and video traffic \cite{americas}.

Despite this in depth level of security design there are still areas which need to be addressed in the 5G security model. This includes new attack surfaces introduced by the greater use of cloud and edge computing, as well as the convergence of 5G with traditional networks creating new attack vectors. The approach taken in this paper by applying anomaly detection is the attempt to detect all traffic that is undesirable in the network, this means that malicious traffic that impacts both the network and potential end users can be detected earlier to minimise adverse effects. Malicious attacks can be generalised into two categories zero-day attacks and day-one attacks \cite{americas}. Zero-day attacks are threats that do not have an existing fingerprint or signature, day-one attacks are threats that have a signature or fingerprint and can be effectively mitigated. The end goal of anomaly detection is to provide a faster and more proactive response to previously unseen (zero day) threats and appropriate mitigation.

\subsubsection{Future Security Concerns}
In addition to the new services and capabilities that 5G networks will provide to users, 5G will bring a host of new security concerns and considerations. These security challenges for 5G can be broken down into four main categories, the management of IoT/V2X/M2M (Vehicle to X, Machine to Machine), distributed architectures, virtualisation and multiple technologies \cite{cisco1}. IoT devices themselves are cheap devices designed for a specific use and security is usually an afterthought, most of these devices do not even have their own IP stack, let alone an inbuilt security system. Communication to an end user from the IoT devices is also another cause for concern due to peer-to-peer communication having no controller between parties, this is a major threat surface. 

Distributed architecture relates to the separation of control and user plane. For example traditionally a packet core network is comprised of all hardware components located in a data centre and these components have known parameters and interfaces. However with 5G, core components can be deployed on the edge and due to the nature of 5G being a cloud native architecture these components are also now on cloud servers. This creates new threat surfaces due to the added difficulty of having to manage a distributed packet core. The heavy use of virtualisation means that communication between parties is web based and accomplished through the use of API's (Application Programming Interfaces), these API's do not have set interfaces and defined common protocols in comparison to an LTE network, therefore this creates an additional threat surface. 5G also becomes another network to manage in the heterogeneous mix of networks currently in operation. Security processes also need to address securing the connectivity elements between 3G, LTE and 5G networks.

A high level view of the 5G threat landscape is shown below in Figure 2, highlighting these security challenges and network segments that are at risk. Threats can be broken into categories based on which parts of the network they are impacting \cite{americas}:

\begin{itemize}
    \item \textbf{User Equipment Threats}: Mobile botnets can to launch DDoS attacks on multiple network levels impacting 5G infrastructure, web servers and user equipment. The goal is to bring services offline.
    \item \textbf{Cloud Radio Access Network Threats}: Rogue base station threat to facilitate as a MITM(Man in the Middle) attack, this attack can compromise user information, tamper with information, track users or cause DoS attacks. Exploit 5G/LTE inter-networking and launch downgrade attack.
    \item \textbf{Core Network Threats}: Vulnerable to IP (Internet Protocol) based attacks from the internet, a botnet can launch user plane and control plane attacks to degrade or put critical core infrastructure offline.
    \item \textbf{Network Slicing Threats}: virtualisation based threats due to the reliance on the security of the hypervisor. Need to ensure isolation of slice functions and resources from other slices, also authentication from user equipment operating on a slice.
    \item \textbf{SDN (Software Defined Networking) Threats}: Separation of control and user plane allows a malicious user to attack the link between control and user plane, a DoS (Denial of Service) attack could be performed or control could be gained over network elements.
\end{itemize}

\hspace*{-0.5cm} 
\makebox[\textwidth][l]{\includegraphics[scale=0.3,width= 0.55\textwidth]{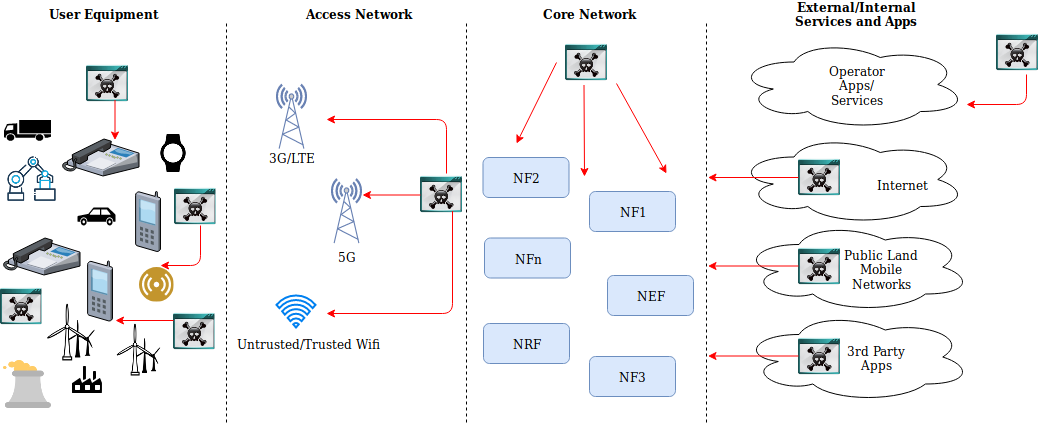}}
\textit{Fig. 2: End-to-end 5G Network Threats overview  \newline Core Network Elements: Network Function (NF, NFn), Network Exposure Function (NEF), Network Repository Function (NRF)}

Managing end-to-end encrypted traffic is another consideration in the evolving 5G threat landscape, as traffic visibility becomes limited inside 5G networks due to encryption and web services further encrypt their traffic. Encrypted traffic has increased by more than 90\% year by year, with a predicted amount of 80\% of all web traffic to be encrypted in 2019 \cite{cisco2}. The encryption of network traffic allows much greater levels of privacy and security, however this same encryption hinders network operators visibility of traffic and therefore the ability to determine if this traffic is malicious or benign. Mobile, cloud and web applications depend on well implemented encryption mechanisms, utilising keys and certificates to verify trust. The advantages of encryption are also its disadvantages as malicious users can employ encryption to evade detection and secure their malicious activities. 

The issue then in terms of security is that the majority of organisations do not have the tools or solutions to manage potentially malicious encrypted traffic and systems are not in place that have the ability to effectively detect malicious encrypted traffic without performance impacts to the network \cite{cisco2}. Traditional techniques such as deep packet inspection become more difficult to perform as traffic would need to be decrypted at some point in the network, analysed and then re-encrypted, this would be a resource and time intensive process. Instead both encrypted and unencrypted traffic can be analysed with flow related statistics. A network flow can be defined as a stream of traffic with a common set of identifiers \cite{netford}. Analysing flow statistics with machine learning will allow the detection of malware in encrypted and unencrypted traffic, without the need to decrypt and re-encrypt every flow.

\subsubsection{5G SDS Implementation Architecture} 
5G networks and their major elements such as the C-RAN (Cloud Radio Access Network) and core network are virtualised, therefore completely defined through software. A similar approach can be taken for implementing an automated security system through SDS. Figure 3. below shows a possible implementation of an SDS system in a 5G network. A copy of a sufficient amount of traffic from both the backhaul link and from the core network link can be analysed to provide end-to-end network anomaly detection. A copy of data is taken for analysis and to build profiles of defining benign and anomalous traffic for the model, also by copying data there will be no impacts on network performance while the model analyzes the data. The data is then pre-processed to be in a form appropriate for the machine learning model and analyzed for anomalies, any identified anomalies are then stored in the policy manager database with the corresponding traffic features. These policies are then sent to a VNF (Virtual Network Function) manager which can then update the appropriate IDS (Intrusion Detection System) module in the core network. Based on the time it takes for the model to process the data, set schedules can be defined for running the model to ensure policies in the IDS Module are kept up to date and to further enhance learning of the machine learning model. The key benefits are the ability to automate the detection, database updates and appropriate action of any malicious flows.

Figure 4. below displays how this SDS system can also be deployed on specific network slices to monitor traffic flows and build benign and anomalous traffic profiles based on the required specifications for that slice. The layout of the diagram focuses on the separation of CP (Control Plane) and UP (User Plane), with UP's residing either in the network core or in the C-RAN, UP's can reside in the C-RAN if being closer to the edge is required for latency reasons, CP's reside in the network core to centralise control of the network. C-RAN elements are distributed including the vBBU's (virtualised Base Band Units), MEC (Mobile Edge Computing) applications and UP's. The coloured lines indicate the logical connections between the SDS system and various network components, slice data is accessed both from the first DC (Data Center) to monitor backhaul traffic from the C-RAN and also from the distribution of network slices within the core network. The key benefit of a SDS system is that it can be deployed in different parts of the network efficiently and with low cost. By developing software defined 5G security tools in a slice based approach, anomaly patterns can be defined per slice. One example of this is training the model to identify infiltration attacks for small IoT devices operating on one network slice that could be potentially used in botnets for DDoS attacks. Depending on operator requirements each SDS system is customisable to their needs.

\vspace{0.1cm}
\hspace*{-0.3cm}
\makebox[\textwidth][l]{\includegraphics[scale=0.3,width= 0.5\textwidth]{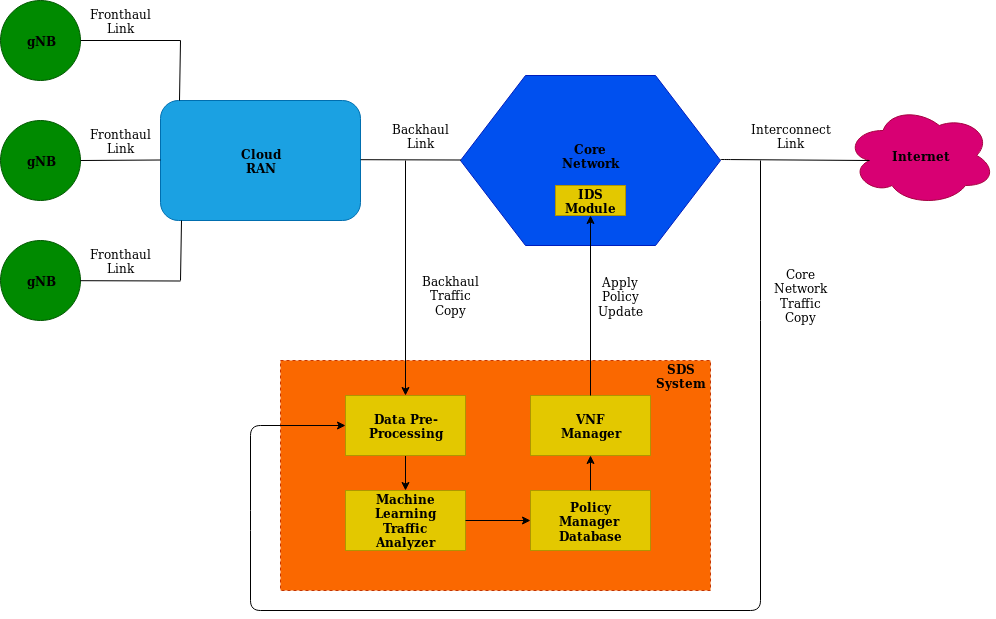}}
\textit{Fig. 3: 5G Network with SDS System}

\hspace*{-0.4cm}
\makebox[\textwidth][l]{\includegraphics[scale=0.3,width= 0.5\textwidth]{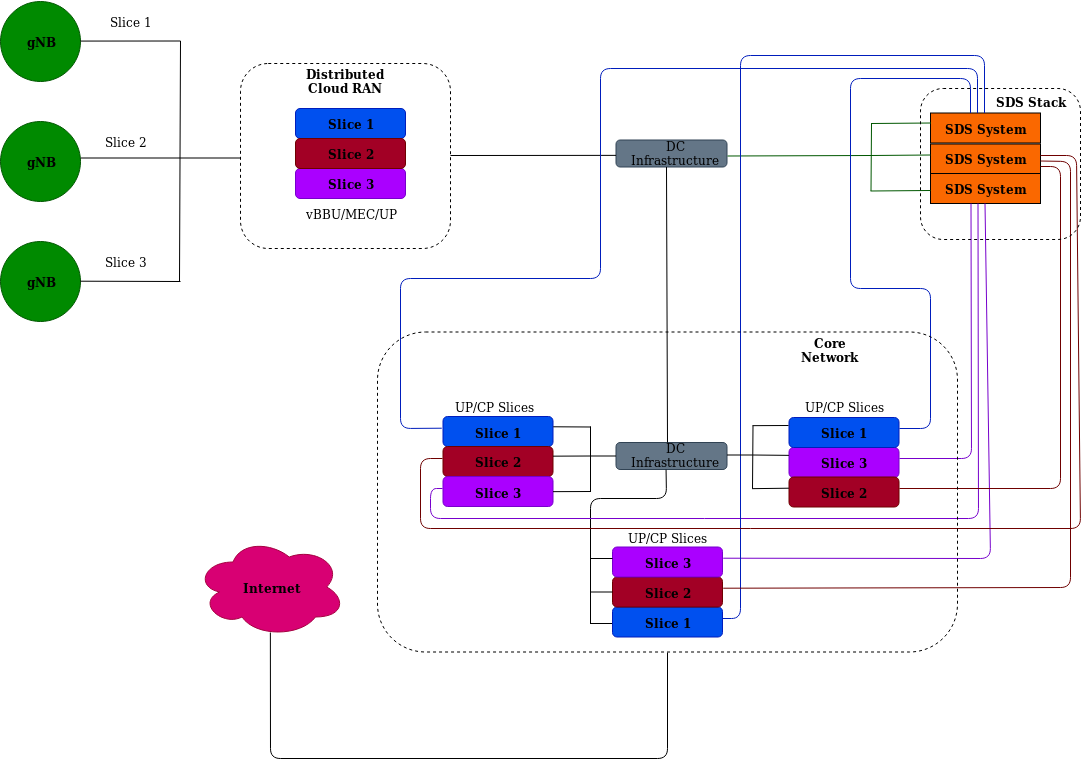}}
\textit{Fig. 4: SDS Application to Network Slices}

\section{Convolutional Neural Network Anomaly Detection}
Deep learning is an area of machine learning which involves the design of multi-layered neural networks, which are essentially mathematically based neuron-like structures that use many variables to solve a complex equation. To develop a neural network for classification of text or images requires significant amount of architectural engineering to obtain a network that is best suited to the provided data set and has a sufficient level of accuracy \cite{learning trans} . This section will therefore explore current advances in the area of machine learning based anomaly detection and then investigate how  techniques such as autoML and NAS can optimise model design to allow the design of CNN architectures that are both scalable and highly optimised for the type of data they are training on.

\subsubsection{Current Advances}
Network intrusion detection relates to the issue of monitoring and differentiating normal network flows from abnormal flows which can compromise the security of a system. Both governments and organisations invest heavily to find a reliable solution to protect their information assets and resources from malicious access, this has brought intrusion detection systems to the forefront of the cyber security landscape \cite{enhanced anom}. As proposed by Denning \cite{denning} the idea of developing intrusion detection systems that employ machine learning techniques is to identify abnormal usage patterns and abnormal traffic which may signal an attempted intrusion of the network. This notion led to the creation of a new type of IDS based on learning algorithms rather than manually updating signatures from previously identified intrusions. Over the last three decades various machine learning techniques have been applied in a conventional approach for developing network anomaly detection models. These approaches employed supervised, unsupervised and semi-supervised learning algorithms to propose a solution for anomaly detection \cite{enhanced anom}. 

Therefore anomaly detection is not a new area of study in machine learning applications and current research has explored a variety of machine learning based applications. However some common issues arise such as low accuracy levels due to sub-optimal model design, unrealistically high accuracy levels due to a lack of model generalisation and over fitting, and also the use of out of date and simplistic data sets. As shown in \cite{dl anom} accuracy over 99\% is achieved using a multi-layered neural network, however the data set used is the KDD99 dataset, a data set which is 20 years old and does not represent current dynamic network environments.

Anomaly detection itself can be most easily modelled as a classification problem in supervised learning \cite{enhanced anom}. Supervised learning means that labeled data is used to train the anomaly detection model. The goal of this type of training is to classify the test data as anomalous or normal on the basis of a specific set of features. In this paper the anomaly detection problem will be approached from a supervised learning perspective and use a CNN architecture designed using NAS to attempt to optimise the highest possible accuracy levels.

Effective model design requires a significant degree of architectural engineering \cite{learning trans}, such as \cite{emp study anom} demonstrates that the design of basic CNN's where extra layers are just added for testing purposes does not improve accuracy, giving sub-optimal results at under 80\% detection rate. \cite{deep and machine} demonstrate the effectiveness of up and down sampling on data to equalise volumes of anomaly and benign data, achieving a detection rate of 99.99\% using random forest and 99.30\% using three layered deep neural networks, these very high results  are unlikely to represent real world detection levels and give the impression of an over fitted model and a lack of generalisation. Effective classification of both benign and anomalous traffic is also an issue, in most cases models can identify labelled benign traffic with very high (99-100\%) accuracy, however determining anomalous traffic can be more difficult, as shown in \cite{ad iot} where the random forest algorithm is applied to the UNSW-NB15 dataset,  benign traffic was classified at 99\% accuracy, however anomalous traffic was classified at 82\%, this means that 18\% of anomalous traffic was essentially undetected.

The approach of this paper attempts to rectify and address some of these common issues. This is done in two main ways by selecting the most up to date IDS data set, the CICIDS2018 which simulates a real world environment and is explained in detail further on. And secondly by using a CNN model based on NAS, which has achieved some of the highest accuracy levels in the ImageNet data set and uses a controller to autonomously optimise parameters for the model. By taking this approach the most optimal model can be generated for a specific data set.

\subsubsection{AutoML \& NAS Implementation}
Neural architecture search brings automation to the design of neural network models, this allows the most optimised model designs to be computed without the tedious process of physically designing, testing and adjusting models. This cutting edge technique in neural network design has led to the rise of a number of automated machine learning platforms. In this paper Google's autoML Vision and Vision Edge platforms will be utilised for model design, training, validation and testing. The underlying architecture which enables these platforms is NASNet (Neural Architecture Search Network) and MNasNet (Mobile Neural Architecture Search Network).

Neural architecture search can be defined as a gradient-based method for finding optimised architectures. The structure and connectivity of a neural network can be specified by a variable length string. Therefore it becomes possible to use a RNN (Recurrent Neural Network) as shown in Figure 5. to generate this string \cite{nas}. The network specified by the string is known as the child network and training the real data set with the child network will result in progressive accuracy increases on the test data set. This accuracy can be used as the reward signal to compute the policy gradient to update the controller. Therefore in the next iteration the controller will give a higher probability to architectures that receive a higher accuracy \cite{nas}. Put simply this means the controller can learn to improve its search over time and optimise placement of layers and blocks of the neural network \cite{seif automl}.

\includegraphics[scale=0.3,width= 0.5\textwidth]{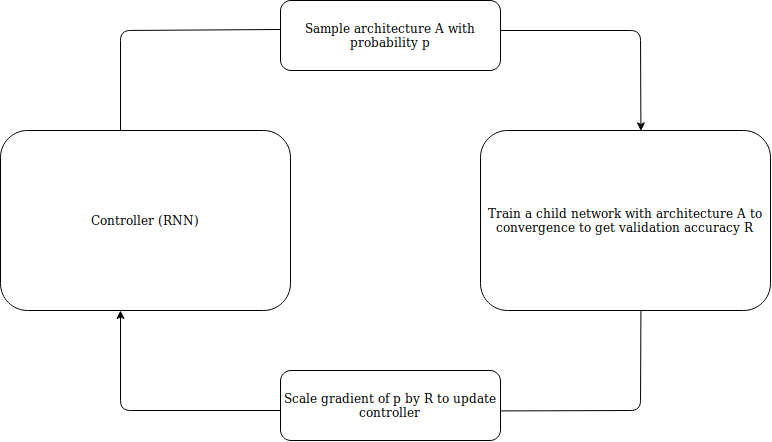}
\textit{Fig. 5: RNN Controller}

In terms of implementation neural architecture search uses the controller to generate a set of architectural hyperparameters of the network. In the case of a CNN it can predict filter height, filter width, stride height, stride width and a number of filters per layer \cite{nas}. This process is then repeated until the number of layers exceeds a certain value. 

This issue with NAS is applying it to a very large data set is extremely computationally intensive. Therefore this technique is applied to a sample of the data set \cite{learning trans}. The NAS search space is defined so that the complexity of the architecture is independent of the depth of the network and the size of input images. It achieves this by breaking down all CNN's in the search space into cells with identical structure but different weights as shown in Figure 6 \cite{learning trans}. Therefore searching for the most optimal architecture can be reduced to searching for the best cell architecture. By searching for each specific cell architecture, speed is greatly increased and the cell is more likely to have better generalisation. Based on this individual cell training approach, networks can be optimised for speed or accuracy depending on the search space size. This allows the neural network to achieve a very high level of accuracy on the ImageNet validation data set at 82.7\% top 1 accuracy \cite{papers code}. ImageNet is the largest database for labelled images containing over 14 million images and has widespread use in providing a benchmark for determining the performance of different CNN models \cite{imagenet}.  
\hspace{-0.2cm}
\includegraphics[scale=0.3,width= 0.5\textwidth]{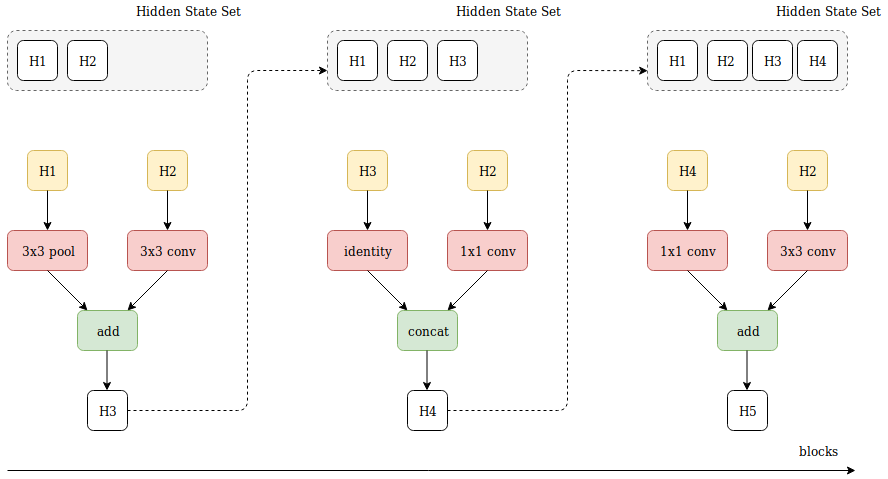}
\textit{Fig. 6: NAS Search Space Block Generation}

MnasNet extends the NAS search space concept by implementing factorised hierarchical search space \cite{mnas}. The factorised hierarchical search space encourages additional layer diversity throughout the network and balances the size of the total search space. This approach brings more flexibility into NAS as models can be designed to balance speed vs. accuracy. So far this approach has the biggest advantage of speed. On the ImageNet data set the MNasNet architecture achieved 75.2\% top 1 accuracy which in comparison to traditional mobile neural network architectures is 1.8 times faster than MobileNetV2 \cite{mobilenet} and 0.5\% higher accuracy. In comparison to NASNet results were 7.5\% lower accuracy, however 2.3 times faster in processing images within the architecture \cite{mnas}.

In the results section NASNet and MNasNET will be compared with tests conducted for 24 hours and 3 hours respectively to assess the differences in results. Latency and computational power is also a primary concern for implementation purposes in this case. By optimising a neural network that can still achieve a high level of accuracy, low latency when training and also can be run on devices such as a modern day smart phone, this will allow much more flexibility for deployment in a 5G network.

\section{Anomaly Detection Data Set}
\subsubsection{Data Set Environment Overview}

Anomaly detection is one of the most promising areas of research in detecting novel attacks. However its adoption to real world applications is hindered due to system complexity requiring a large amount of testing, tuning and evaluation. Therefore for research purposes a simulated system can be designed with a comprehensive set of intrusions and abnormal behaviour mixed in with normal traffic for anomaly detection analysis.
As network behaviours and malware are changing it becomes necessary to have an environment that more accurately simulates a real world scenario. The data that can then be captured from the system is dynamic and provides more meaningful and realistic insight into benign and anomalous network traffic behaviour. Unfortunately traditional IDS data sets were not designed in this way, for example the KDD CUP99 data set or the ADFA-IDS data set were created in a testing environment that was only comprised of single LAN links and one attacking and one defending system, this approach represents a static environment and provides sub-optimal and less realistic results \cite{azsecure} \cite{kdd}.

The IDS-2018 data set from the Canadian Institute of Cybersecurity is a data set derived from a simulated environment that attempts to address these shortcomings \cite{cic}. The main objective of this data set is to use a systematic approach to generate a diverse and comprehensive benchmark data set for intrusion detection based on the creation of benign traffic and malicious traffic profiles. The environment itself is comprised of 50 attacking machines on a victim organisation with 5 departments which includes 420 machines and 30 servers. The data set takes packet captures of network traffic and system logs of each machine, as well as the extraction of 80 network features organised as flows. Figure 7. below shows the overall network topology which is a common LAN network on an AWS (Amazon Web Services) cloud platform. 6 subnets are installed labelled as Dep1 to Dep5 and Servers.
Dep1 to Dep4 machines have Windows 8/10 OS's, Dep5 has all Linux machines running Ubuntu, Servers has different MS Windows servers such as App servers, active directory and email. The attacker network has Windows 8/10 machines and Ubuntu machines.

\includegraphics[scale=0.3,width= 0.4\textwidth]{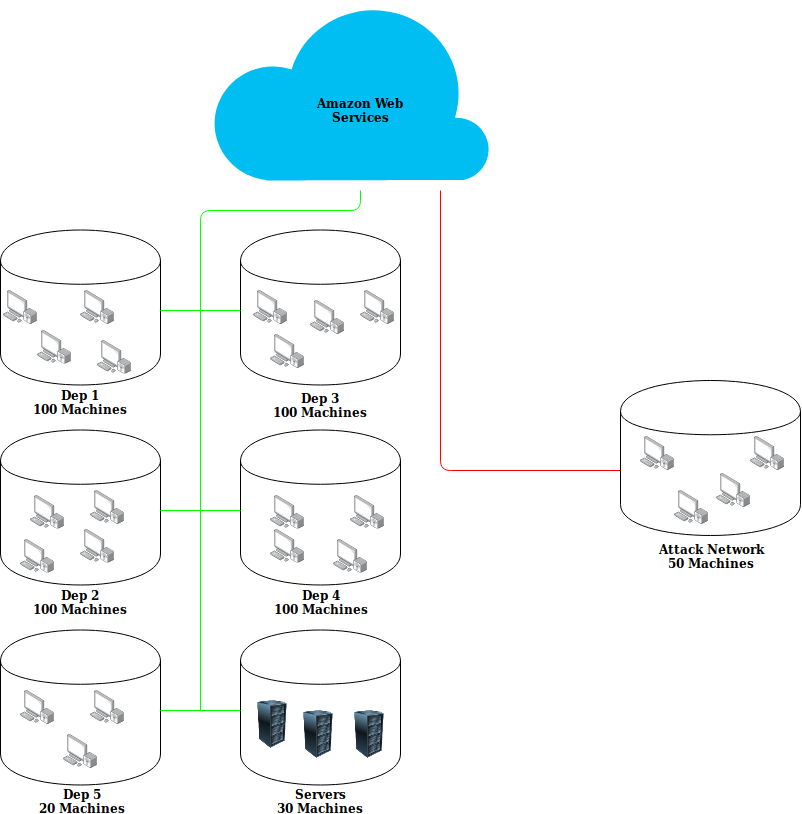}
\textit{\newline Fig. 7: CICIDS2018 Network Topology}

\subsubsection{Network Profiles \& Features}
Protocols simulated in the environment are: HTTPS (HyperText Transfer Protocol Secure), HTTP (HyperText Transfer Protocol), SMTP (Simple Mail Transfer Protocol), POP3 (Post Office Protocol 3), IMAP (Internet Message Access Protocol), SSH (Secure Shell), FTP (File Transfer Protocol). Traffic types are broken into two profiles, either a B-profile (benign traffic) or M-profile (malicious traffic). The types of traffic within these profiles is explained in further detail below.

\textbf{B-Profile:} Describes normal traffic types simulated through a number of machine learning algorithms with different network protocols \cite{cic}:
\begin{itemize} 
  \item Emulates the behaviour of users by utilising various machine learning statistical analysis techniques such as K-Means, Random Forest, SVM and J48. 
  \item Network features collected include packet size of protocol, number of packets per flow, various patterns in payload, size of payload and request time distribution of a protocol. 
\end{itemize}

The specific attacks used in the M-Profile are common attacks used by malicious actors as well as penetration testers. They cover a wide variety of scenarios from network based attacks, different forms of HTTP DoS and DDoS, brute force attacks, web based attacks and widespread vulnerabilities. They also cover aspects of the OWASP top 10 2019 including injection based attacks from SQL, broken authentication due to poor password management allowing easier brute force attacks and security misconfigurations which allow vulnerabilities such as heartbleed due to unpatched systems \cite{sucuri}.

\textbf{M-Profile:} Describes the attack scenario for anomalous traffic, six different attack scenarios are simulated \cite{cic}:
\begin{itemize}
  \item Internal network infiltration - exploits application vulnerability by sending malicious files via email. Metasploit framework is utilised for exploitation allowing a backdoor to be executed on the victim's PC.
  \item HTTP DoS - Slowloris, LOIC and HOIC which cause denial of service are used, these tools are able to make web servers inaccessible. Slowloris can do this with just one machine and is most effective against Apache servers \cite{medium}. Apache servers are the second most common web servers on the internet accounting for 26.73\% of web servers \cite{netcraft}.
  \item Web app attacks - Web application based attacks tested using the Damn Vulnerable Web App (DVWA) for SQL injection, command injection and unrestricted file upload.
  \item Brute force attacks - Use a dictionary brute force attack containing 90 million words against main servers to attempt to acquire SSH and MySQL account information.
  \item Last updated attacks - Well known vulnerabilities that can affect thousands of devices under certain conditions and if they are running older, outdated versions of software. Heartleech will be used in this environment, it is used to scan systems vulnerable to the Heartbleed bug, once systems are found they can then be exploited and data can be exfiltrated.
\end{itemize}

To define the features from these profiles, initial raw packet captures are converted to network flows for easier analysis. Using CICFlowMeter bidirectional flows are generated where the first packet determines the forward (source to destination) and backward (destination to source) directions. Therefore from the 83 statistical features gathered from the flows such as duration, number of packets, number of bytes, length of packets, these are calculated separately for both forward and reverse directions. For TCP flows they are terminated upon connection teardown (once a FIN packet is received) and UDP flows are terminated by a flow timeout.


\hspace*{-0.75cm}  
\includegraphics[scale=0.3,width= 0.65\textwidth]{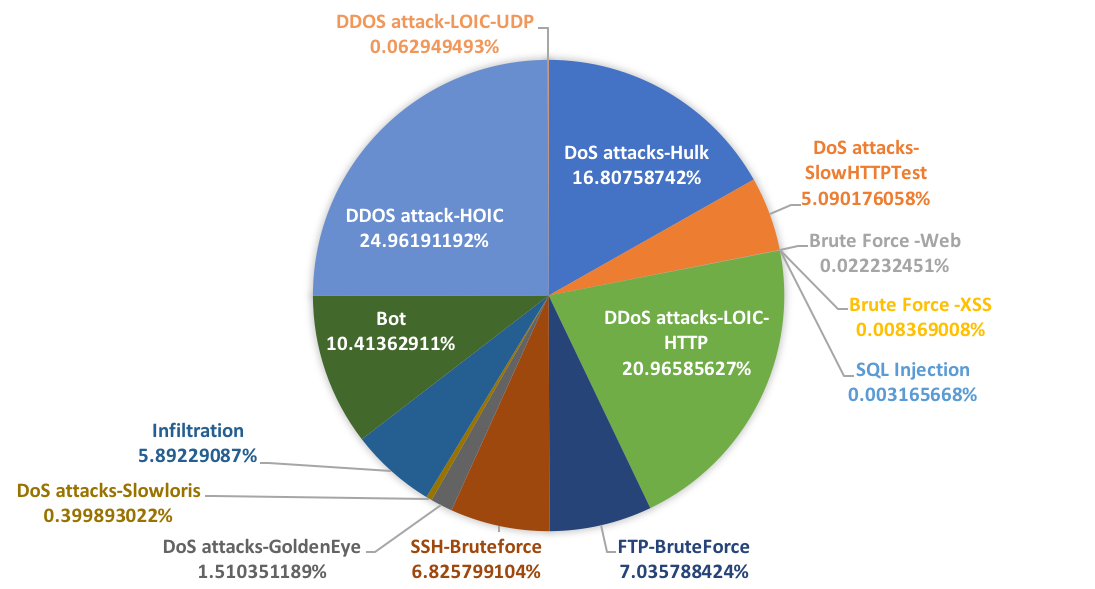}
\textit{\newline Fig 8. Pie Chart of Malicious Traffic Type Volumes}

\includegraphics[scale=0.3,width= 0.4\textwidth]{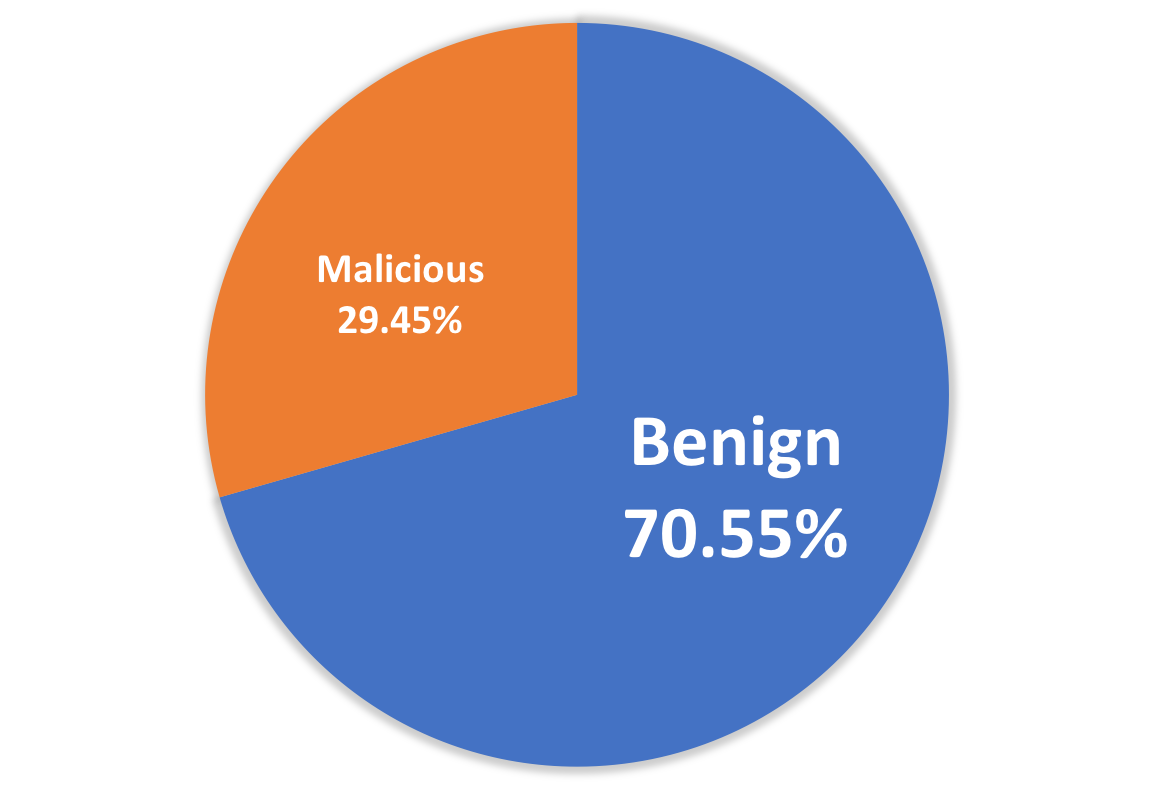}
\textit{\newline Fig 9. Pie Chart of all Traffic Volumes}

This paper will break down all labelled network flows into two streams for analysis, anomalous and benign. Benign is comprised of all traffic described in the B-Profile and anomalous is all traffic described in the M-Profile. Different attacks occur at different days out of a total of 10 days or 240 hours, these attacks are dispersed randomly within benign traffic. In total there are 2748235 anomaly flows and 6584535 benign flows giving a total of 9332770 flows in the data set. This is a split of 70.55\% benign traffic and 29.45\% anomaly traffic. The two pie charts below in Figure 8. and Figure 9. show the breakdown of traffic volumes in the data set.

\section{Inter-Arrival Time \& Feature Selection}
IAT (Inter-Arrival Time) can be defined as the average frames, packets or flows that arrive at a host over a certain time period \cite{interarrival}. By examining this feature and other statistical forms of IAT such as the mean, minimum, maximum and standard deviation of IAT of a network flow, benign traffic can be modelled to conform to the Weibull distribution. By modelling benign traffic to the Weibull distribution, anomalous traffic can therefore be identified as it will cause irregularities and deviations in the distribution \cite{flow interarrival}. This correlation is identifiable across packets, flows and sessions for both TCP (Transmission Control Protocol) and UDP (User Datagram Protocol) transport protocols in internet traffic \cite{weibull}. Therefore these IAT network flow features can be indicative of the difference in benign and anomalous flows.

Current studies demonstrate the Weibull distribution modelled to internet traffic by using traffic traces captured from the WAND Research Group \cite{wand}. 24 hours of traffic monitoring from an ISP has captured data from wireless hotspots, DSL and ethernet connectivity in an urban environment \cite{weibull}. The data captured shows the conformance for packets, flows and sessions as they decrease from unity (value of 1) to the Weibull distribution.

Extending this concept to focus on network flows, \cite{flow interarrival} has demonstrated that despite the variety of networks in size, number of users, applications and loads, the IAT's of benign TCP flows also conform to the Weibull distribution and specific irregularities in these flows will cause deviations in traffic. Multiple data sets were collected with differing bandwidth, size and applications to verify this conformance. Data sets that were tested in \cite{flow interarrival} were:

\begin{itemize}
    \item \textbf{1. MAWI3(Measurement and Analysis on the WIDE Internet)}: June 2012, 1.4 million flows, captured from a 150Mbps trans-pacific backbone link between Japan and USA.
    \item \textbf{2. SUT(Sharif University of Technology)}: June 2012, 2.4 million flows,  captured from internet gateway of SUT campus.
    \item \textbf{3. MCO}: February 2011, 2.3 million flows, captured from an internet gateway of a medium business company.
    \item \textbf{4. NUST1(National University of Sciences and Technology)}: March 2009, 2.2 million flows, Captured from an endpoint router located in NUST, Pakistan.
    \item \textbf{5. ISP\_NUST}: captured from an edge router of a medium sized ISP and merged with attack flows generated in NUST.
\end{itemize}
When analysing the traffic flows from the above data sets, \cite{flow interarrival} shows that the deviation in the Weibull distribution is visible when comparing all flows to benign flows.

Specific attack injections into the ISP\_NUST data set have also been analysed in terms of detection rate. The attack injection was for a SYN flood attack, a type of DoS attack that consists of a high volume of SYN packets with a very small inter-arrival times \cite{flow interarrival}. This drastic change in inter-arrival time causes irregularities in the Weibull distribution and allowed detection of attacks. \cite{flow interarrival} shows that a 98.8\% accuracy rate was achieved with a 4.8\% false alarm rate. This demonstrates the high amount of variance that some common types of malicious attacks can have on flow inter-arrival time. This notion forms the basis for the feature selection decision from the CICIDS2018 data set and these assumptions will be verified in the results section.

By considering these past studies in inter-arrival traffic flow behaviour these concepts can be extended to current day machine learning models to provide clearly defined labelled data on classifying between anomalous and benign traffic flows. Feature selection therefore involved a two part selection process. The first part is the selection of standard features that provide basic information on the flow. The second part involves selecting a limited number of features that demonstrate clear differences in values between a benign and anomalous flow. As demonstrated previously, IAT flow data can be proposed as a strong candidate and more specifically statistical variations of IAT flow data can be used to further analyse these correlations. This decision to limit feature selection is to provide the machine learning model with clean data and to remove excess noise in the data that is not meaningful in correlating the relationship between anomalous and benign flows. By doing this a more efficient model can be designed, with higher accuracy and faster speed. 20 features along with an additional label column to classify each flow type have therefore been selected and these are:

\begin{itemize}
\item \textbf{Basic Flow Features}: Destination Port, Protocol, Flow Duration, Total Forward Packets, Total Backward Packets, Flow Pkts/s

\item \textbf{IAT Statistical Metadata}: Flow IAT Mean, Flow IAT Standard Deviation, Flow IAT Maximum, Flow IAT Minimum, Flow IAT Total, Forward IAT Mean, Forward IAT Standard Deviation, Forward IAT Max, Forward IAT Min, Backward IAT Total, Backward IAT Mean, Backward IAT Standard Deviation, Backward IAT Max, Backward IAT Min

\end{itemize}

\subsubsection{Pre-Processing Data Set}
Data set pre-processing involves transforming the input data into the correct form suitable for the CNN, which in this case is a 100x100x3 image. The defined 20 features from the data set are extracted in CSV file format. CSV inputs are reshaped into RGB images of 100 x 100 x 3 size, any additional left over data under this size is discarded as all images for the CNN are required to be of the same input size. This image size was chosen due to providing a good volume of sample images for the amount of data available (over 1000 sample images). In general the trade offs between using a higher compared to a lower resolution image is that a higher resolution image will contain finer details when processed by the neural network, however this will take longer for both training and testing phases. A lower resolution image will provide less details, but more global feature representations and the neural network will be able to train and test the data at a faster rate. For this paper autoML samples and augments all images to 224 x 224 x 3 input image size, therefore there are only two considerations, firstly the volume of images is above 1000 and that sufficient feature details are captured. Two examples are shown below in Figure 10 and Figure 11 of what an anomaly image looks like in comparison to a benign image. Graphically, anomaly images are random and noisy, whereas benign images are more regular and contain some identifiable patterns.

\begin{figure}[htbp]
    \centering
    \subfloat[Anomaly Image 1]{{\includegraphics[scale=0.3,width=2.8cm]{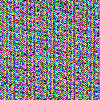} }}
    \qquad
    \subfloat[Anomaly Image 2]{{\includegraphics[scale=0.3,width=2.8cm]{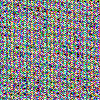} }}
    \\
    \vspace{3mm}
    \textit{Fig. 10: Anomaly Image Example}
    \\
    \centering
    \subfloat[Benign Image 1]{{\includegraphics[scale=0.3,width=2.8cm]{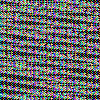} }}
    \qquad
    \subfloat[Benign Image 2]{{\includegraphics[scale=0.3,width=2.8cm]{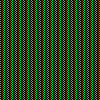} }}
    \\
    \vspace{3mm}
    \textit{Fig. 11: Benign Image Example}
\end{figure}

\section{Anomaly Detection Results}
This section presents the results of the implemented autoML Vision model on the pre-processed image data. To maximise the accuracy of the model both a 24 hour NASnet simulation and a 3 hour MNasNet simulation are run to compare the attained results.
Now that the data set has been pre-processed into images, these images can now be uploaded to Google AutoML Vision. The images are organised by folder structure for benign and anomaly images, these images are then uploaded to Google Cloud Bucket storage along with a CSV file to map output file paths to the correct label. AutoML Vision reshapes the input image sizes into the models expected input size of 224 x 224 x 3. See appendix A for model layout.
With a training time of 3 hours and 1433 100 x 100 x 3 images split into 925 benign images and 508 anomaly images. 141 test images achieved an average precision of 97.6\%, a max precision of 98.582\% and a max recall of 98.582\% for the entire model. As the model is developed on the MNasNet Edge platform it is more optimised for speed in comparison to traditional NAS models and does not require huge amount of compute power and multiple days for training. Training the model for higher accuracy yields a process time of 105ms per image for a Pixel 1 mobile phone. 

Running the model on the 24 hour test yielded similar results, except average precision was slightly higher at 99.2\%, this is due to the area under the recall precision curve being greater as shown in Figure 12. compared to Figure 13. This means the model is optimised to its greatest potential with the data provided. Training with additional data sets and more data will only make the model have a higher level of performance. In terms of real world implementation there will always be a trade off between accuracy and speed. However in this case the small loss in accuracy for a much larger gain in performance is desirable. This means that resource usage can be minimised, threats can be detected sooner and subsequent additional training and refining of the model can be completed at a faster rate.

\makebox[\textwidth][l]{\includegraphics[scale=0.3, width=0.50\textwidth]{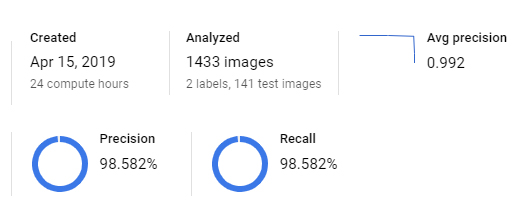}}
\textit{Fig. 12: 24 Hour Test Results}

\hspace*{-0.4cm}     
\makebox[\textwidth][l]{\includegraphics[scale=0.3, width=0.50\textwidth]{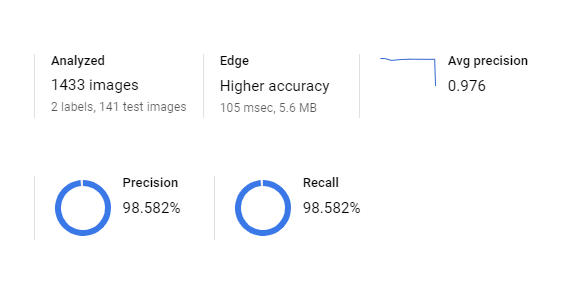}}
\textit{Fig. 13: 3 Hour Test Results}

The most common metrics are used to determine relevancy of results including  precision, recall, f1 score and precision recall curve. Precision and recall are determined from the following stats:

\begin{itemize}
    \item \textbf{True Positive (TP)}: An anomaly image is classified by the model as an anomaly the result is a True Positive.
    \item \textbf{False Positive (FP)}: A anomaly image is classified by the model as  benign the result is a False Postive.
    \item \textbf{True Negative (TN)}: A benign image is classified by the model as  benign the result is a True Negative.
    \item \textbf{False Negative (FN)}: A benign image is classified by the model as  an anomaly the result is a False Negative. 
\end{itemize}

Precision can be defined as the percentage of positive predictions that are correct and recall can be defined as what percentage of the positive cases did the classifier detect. Mathematically this can be calculated as:
\vspace{0.2cm}
\[
    Precision = \frac{TP}{TP+FP} \quad Recall = \frac{TP}{TP+FN}
\]

And from these statistics the F1 score which provides the harmonic mean of the precision and recall can also be calculated:
\vspace{0.2cm}
\[
    F1\ score = 2 * \frac{Precision * Recall}{{Precision + Recall}}
\]

The confusion matrix in Figure 14 shows the percentage prediction of when the classifier chose the correct answer, in this case 96.4\% of anomaly images are identified as anomaly images and 3.6\% of anomaly images were incorrectly identified as benign images. For benign images 0\% of benign images were wrongly predicted as anomaly images and 100\% of benign images were correctly identified.

\makebox[\textwidth][l]{\includegraphics[scale=0.3, width=0.50\textwidth]{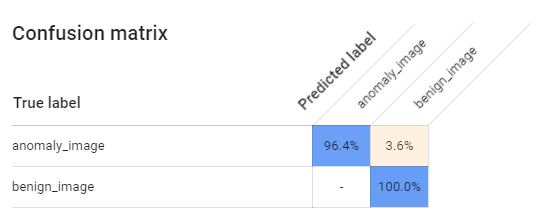}}
\textit{Fig. 14: Confusion Matrix}
\\

Key performance statistics are shown in the table below for easier visibility and comparison. Benign traffic has a recall of 100\% accuracy, but has a  precision of 97.7\% due to classifying a small percentage of anomaly images incorrectly as benign. Anomaly traffic has a precision of 100\% due to all anomaly images detected being classified correctly, but has a recall of 96.4\% due to missing some anomaly images and incorrectly classifying them as benign. 

\textit{Table 1: Performance Statistics}
\\
\begin{tabular}{ |p{1.8cm}||p{1.8cm}||p{1.8cm}||p{1.8cm}|  }
 \hline
 \multicolumn{4}{|c|}{24 Hour \& 3 Hour Results} \\
 \hline
 \textbf{Traffic Type} & \textbf{Precision} & \textbf{Recall} & \textbf{F1 Score}\\
 Benign & 0.977    &1&   0.988\\
 Anomaly& 1  & 0.964 &0.982\\
 Average &0.9885 & 0.982&  0.985\\

 \hline
\end{tabular}
\vspace{0.5cm}

Precision vs. recall is a trade off and this is shown in the below precision recall curves in figure 15. and Figure 16. The score threshold is set at 0.5 to evenly balance these metrics. To simplify these metrics into one number the F1 score can be used, which evenly weights both precision and recall, F1 score values are shown above. The F1 score is an important metric for this model as both recall and precision need to be considered in anomalous traffic detection. In a real world application an IDS needs to minimise the amount of benign traffic flows that are identified as anomalies as much as possible, while still attempting to maximise detection rates of true anomalous traffic. \newline

\makebox[\textwidth][l]{\includegraphics[scale=0.3,width=0.48\textwidth]{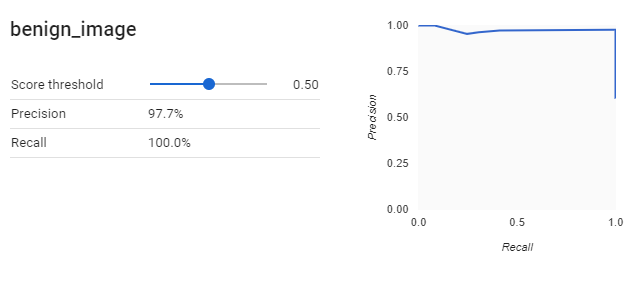}}
\textit{Fig. 15: Benign Data}
\\

\makebox[\textwidth][l]{\includegraphics[scale=0.3, width=0.48\textwidth]{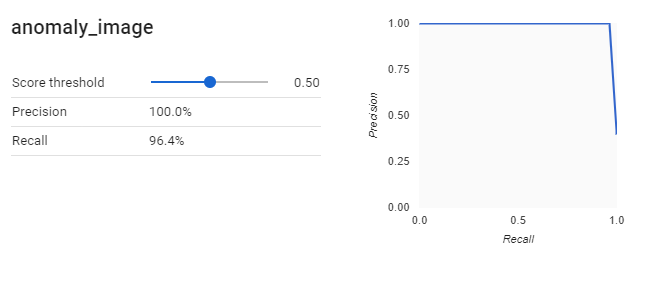}}
\textit{Fig. 16: Anomaly Data}

\section{Discussion \& Future Directions}
There are many areas that need to be considered when attempting to secure 5G networks, as the network is so diverse, security becomes more difficult to implement effectively. The approach of this paper is then to develop an end-to-end monitoring system for traffic that flows through the network, this is not in itself a definitive security solution, just one part of the overall security architecture that is required to secure the network. The model designed with the selected basic and IAT feature set in this paper using autoML has managed to classify all benign traffic flows correctly which is a very good result, however for anomaly traffic flows 96.4\% of traffic was classified correctly, so therefore there is still room for improvement. There was minimal difference in both the MNasNet (3 hour running time) architecture and the NASNet (24 hour running time) architecture, this could be due to the data set size. For further model validation, testing can be conducted with a larger data set, testing can also be conducted with different data sets to ensure a reasonable degree of generalisation in the model and to check for over fitting issues. Finally identifying any visually similar benign and anomaly images and testing with different features to attempt to separate the difference between benign and anomalous images even more could be investigated. The better that an anomaly image and a benign image can be distinguished, the easier it will become to train the model and reduce outlier data and errors. NAS has allowed the creation of an advanced model for a specific data set to be built autonomously and to avoid the tedious process of manual architecture design, the designed NAS model can now be exported into a custom application for further testing and refinement.

Overall the results highlight the effectiveness of machine learning based image detection techniques for network flow analysis. This research could be extended in a number of different directions such as:
\begin{enumerate}
  \item Implementing unsupervised learning techniques to create a semi-supervised learning model, as in reality the majority of network traffic is unlabelled data and pre-processing unlabelled data into clean and organised labelled data is a time consuming process. Extending this concept a benign traffic profile could be designed for a specific network slice using unsupervised learning techniques for general classification and then supervised techniques for additional fine tuning to verify the profile.
  \item Building a database that stores traffic logs, a specific volume of the logs would be taken regularly for training the network, further study could be conducted in how often to re-train the network, with what new incoming data, how long should it take to train and would it be possible to deploy multiple instances of the neural network, so that one instance can train on new data while another instance can be tested on existing data.
  \item Implement a real time traffic monitoring system with a machine learning built profile, this could be designed as an intelligent firewall.
  \item Design an enhanced trust based system to authenticate trust based on predictive flow analysis.
\end{enumerate}

\section{Conclusion}
This paper proposed a novel solution of applying software defined security with machine learning to provide end-to-end protection for 5G networks. The initial project scope has been fulfilled and the approach of converting network flows into images for analysis by a CNN has demonstrated highly accurate results for the data available, especially considering that CNN's are traditionally optimised for real image/photograph detection. The application of a machine learning based SDS system is promising for real world implementation and some of the points outlined above explore this further. However challenges still must be overcome, in terms of managing diverse and complex 5G networks and also managing the large volumes and variations of traffic that will flow through them.

Overall this is only the beginning for machine learning based security applications. The growth in 5G network rollouts, global internet usage, IoT device connectivity and big data analysis will continue to widen and create new attack surfaces. To manage and mitigate these attack surfaces effectively, dynamic and intelligent machine learning security systems that can respond rapidly to threats will be critical.

\clearpage
\appendix[A: NAS Designed CNN Model] 
\makebox[\textwidth][l]{\includegraphics[width= 0.9\textwidth]{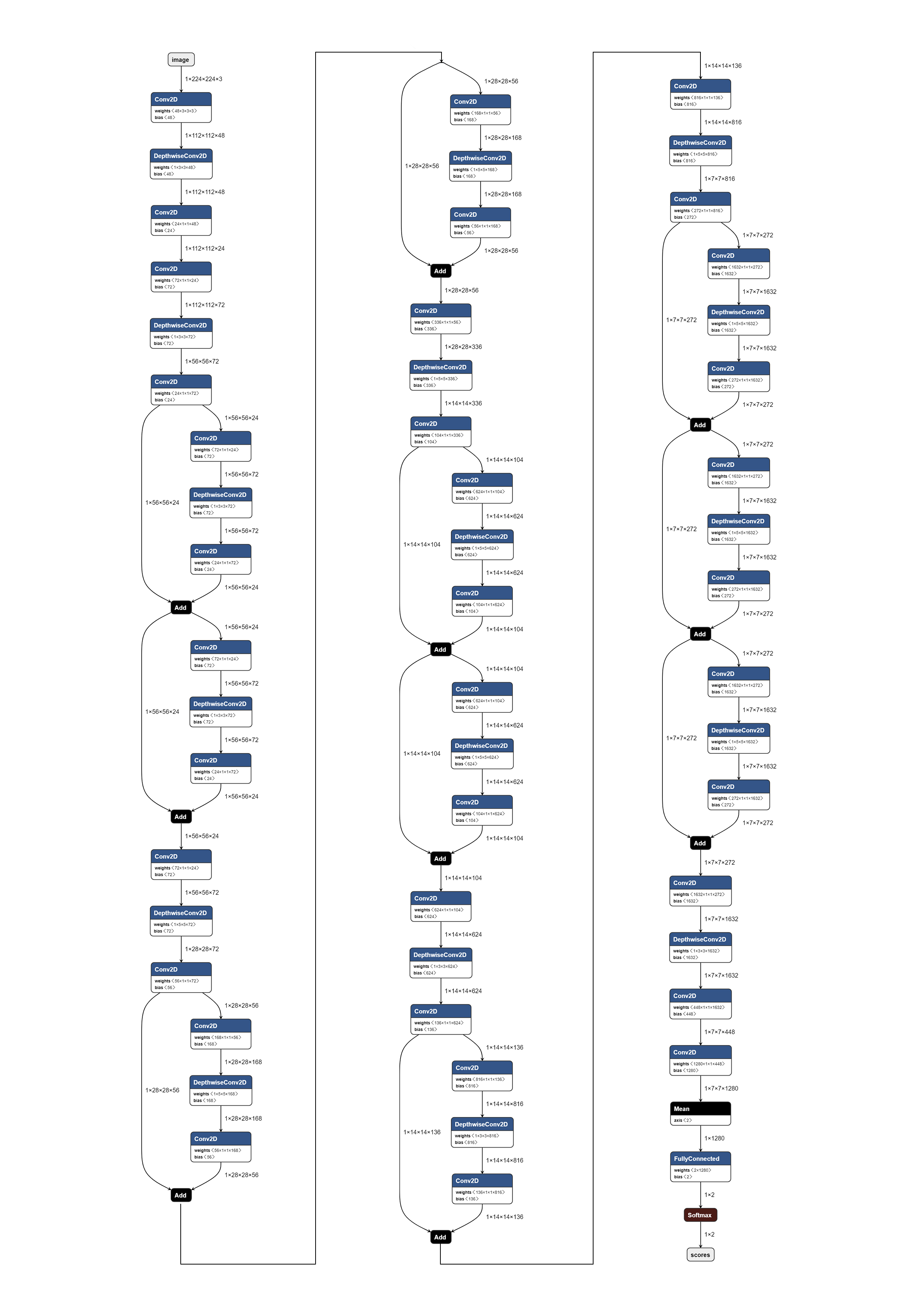}}

%



\ifCLASSOPTIONcompsoc

\ifCLASSOPTIONcaptionsoff
  \newpage
\fi



\bibliographystyle{IEEEtran}
\end{document}